\input{psfig.sty}
\documentstyle[11pt,aas2pp4]{article}

\lefthead{Schachter et al.}
\righthead{NGC 7582}

\begin{document}

\title{NGC 7582: The Prototype Narrow--Line X--ray Galaxy}

\author{Jonathan F.~Schachter, Fabrizio Fiore\altaffilmark{1},
Martin Elvis, Smita Mathur}
\affil{Harvard--Smithsonian Center for Astrophysics,  60 Garden
Street, Cambridge, MA 02138}
\author{Andrew S.~Wilson}
\affil{Astronomy Department, University of Maryland, College Park MD 21218}
\author{Jon A.~Morse}
\affil{Center for Astrophysics and Space Astronomy, University of Colorado,
Campus Box 389, Boulder, CO 80309}
\author{Hisamitsu Awaki}
\affil{Department of Physics, Faculty of Science, 
Kyoto University, Kitashirakawa, Sakyo, Kyoto, 606-01, Japan}
\author{Kazushi Iwasawa}
\affil{Institute of Astronomy, Madingley Road, Cambridge CB3 0HA, United
Kingdom}

% Notice that each of these authors has alternate affiliations,
% which identified by the \altaffilmark after each name.  The
% actual alternate affiliation information is typeset in
% footnotes at the bottom of the first page, and the text itself
% is specified in \altaffiltext commands. 
% There is a separate \altaffiltext for each alternate affiliation
% indicated above.

\bigskip
\bigskip
\bigskip
\bigskip

\centerline{Accepted for publication in The Astrophysical Journal (Letters)}

\altaffiltext{1}{present address: Osservatorio Astronomico di
Roma, Via dell' Osservatorio, I--00140 Monteporzio, Italy}

% The abstract environment prints out the receipt and acceptance
% dates if they are relevant for the journal style.  For the
% aasms style, they will print out as horizontal rules for the
% editorial staff to type on, so long as the author does not
% include \received and \accepted commands.  This should not be
% done, since \received and \accepted dates 
% are not known to the author.

%\date{\tt Version: 5pm 11 May 1997}

\newpage

%%%%%%%%%%%%%%%%%%%%%%%%%%%%%%%%%%%%%%%%%%%%%%%%%%%%%%%%%%%%%%%%
\begin{abstract}

NGC~7582 is a candidate prototype of the Narrow Line X-ray Galaxies
(NLXGs) found in deep X--ray surveys.  An ASCA observation shows the
hard ($>$3 keV) X-ray continuum of NGC~7582 drops 40\% in $\sim$6 ks,
implying an AGN, while the soft band ($<3$ keV) does not drop in
concert with the hard continuum, requiring a separate component. The
X-ray spectrum of NGC~7582 also shows a clear 0.5--2 keV soft ($kT =
0.8^{+0.9}_{-0.3}$ keV or $\Gamma=2.4\pm 0.6$; L$_X = 6 \times
10^{40}$ ergs~s$^{-1}$) low--energy component, in addition to a
heavily absorbed [$N_H = (6\pm 2)\times 10^{22}$~cm$^{-2}$] and
variable 2--10 keV power law [$\Gamma = 0.7^{+0.3}_{-0.4}$; L$_X =
(1.7-2.3) \times 10^{42}$ ergs~s$^{-1}$]. This is one of the flattest
2-10 keV slopes in any AGN observed with ASCA. (The ROSAT HRI image of
NGC~7582 further suggests extent to the SE.)

These observations make it clear that the hard X--ray emission of
NGC~7582, the most `narrow-line' of the NLXGs, is associated with
an AGN. The strong suggestion is that {\em all} NLXGs are
obscured AGNs, as hypothesized to explain the X-ray background
spectral paradox. The separate soft X-ray component makes
NGC~7582 (and by extension other NLXGs) detectable as a ROSAT
source.

\end{abstract}
%%%%%%%%%%%%%%%%%%%%%%%%%%%%%%%%%%%%%%%%%%%%%%%%%%%%%%%%%%%%%%%%
% The different journals have different requirements for
% keywords.  The keywords.apj file, found on aas.org in the
% pubs/aastex-misc directory, contains a list of keywords used
% with the ApJ and Letters.  These are usually assigned by the
% editor, but authors may include them in their manuscripts if
% they wish.

\keywords{galaxies: active, galaxies: Seyfert, X--rays: galaxies,
BL Lacertae objects: general} 

% In the first two sections, you should notice the use of the
% LaTeX \cite command to identify citations.  The citations are
% tied to the reference list via symbolic KEYs.  We have chosen
% the first three characters of the first author's name plus the
% last two numeral of the year of publication.  The corresponding
% reference has a \bibitem command in the reference list below.
%
% Please see the AASTeX manual for a more complete discussion on
% how to make \cite-\bibitem work for you.   

%%%%%%%%%%%%%%%%%%%%%%%%%%%%%%%%%%%%%%%%%%%%%%%%%%%%%%%%%%%%%%%%
\section{Introduction} \label{intro}

Deep X-ray surveys with ROSAT have revealed a rapidly increasing
number of `Narrow-Line X-ray Galaxies' (NLXGs) at moderate
redshifts ($z \sim 0.1-0.5$) and luminosities ($L_X
\sim10^{42}-10^{43}$~ergs~s$^{-1}$, \cite{boy95}, \cite{geo95}).
NLXGs were first discovered in early 2--10 keV sky surveys (e.g.,
{\em Ariel V}, \cite{wa78}; HEAO-1, \cite{picc82}). Since these
NLXGs are highly absorbed ($N_H \sim
10^{22-23.5}$~cm$^{-2}$; \cite{tp89})
at low X-ray energies, their ROSAT counterparts
may be the absorbed AGN population posited (\cite{com95}) to
solve both the hard X--ray background `spectral paradox'
(\cite{bol87}) and the hard X-ray/soft X-ray source counts
discrepancy (\cite{gen95b}, \cite{geo97}).  This picture is not
without problems however: it predicts extremely hard ROSAT
spectra, and ASCA fluxes many times brighter than ROSAT. Yet neither
is found (\cite{geo97}, \cite{rom96}).

The NLXGs may not all be AGN; in high quality optical spectra
half the NLXGs have no broad H$\alpha$ (\cite{boy95}) and so may
be starbursts, rather than AGN.  The brightest NLXG is NGC~7582,
which has linewidths $<$~200~km~s$^{-1}$ (\cite{mor85};
\cite{wn90}) and no broad components out to Brackett-$\gamma$
(\cite{moo90}).  As such, NGC~7582 is the best and the brightest
NLXG candidate for being starburst dominated, and may serve as a
prototype of the new faint NLXG population. X-ray observations
can be critical to determining the presence of an AGN.

Previous X-ray spectra (\cite{mush82}, \cite{wa93}, \cite{tp89})
and variability measurements (\cite{mush82}) of NGC~7582 however,
have been hampered by bright confusing sources: the background
cluster Abell S1101 is 53$^\prime$away (\cite{cha82}), while the
BL~Lac object PKS~2316$-$423 is only $17^\prime$ away
(\cite{craw94}; \cite{st91}).
%
%~\footnote{Although it was labeled a cluster in Stocke et al.,
%the featureless spectrum (Crawford \& Fabian) supports the BL~Lac
%identification (\cite{st95}).}) 
%

In this {\em Letter} we describe ASCA SIS and ROSAT HRI
observations of NGC~7582, separating the contributions of the AGN
and the BL~Lac.  These demonstrate that even this purely 
narrow-line object clearly harbors an AGN; simultaneously these data
suggest an answer to the problems of the soft X-ray flux from
NLXGs.

%%%%%%%%%%%%%%%%%%%%%%%%%%%%%%%%%%%%%%%%%%%%%%%%%%%%%%%%%%%%%%%%
\section{Observations} \label{obs}

%%%%%%%%%%%%%%%%%%%%%%%%%%
\subsection{ASCA SIS Data}
\label{tim_anal}

An ASCA observation in SIS 2 CCD mode (along the diagonal) and
carefully oriented to place both NGC 7582 and PKS~2316$-$423 near
the axis, was performed on 1994 November 11, aquiring 17.0~ks in
SIS0 and 18.2~ks in SIS1 on NGC~7582.  Both sources were clearly
detected.  At 0.1~cts~s$^{-1}$, NGC~7582
dominates the hard (2-10 keV) band by a factor of $\sim$2.3. A
total of 4067 net SIS counts were detected from NGC~7582,
allowing complex spectral fits.  We used the {\tt ascascreen}
default criteria
\footnote{See definitions in Table 5.5.1 of \cite{abcguide}; 
hereafter, ABC Guide. Maximum angular deviation $=$ 0.01; ELV
$>$10; BR\_EARTH $>$20 COR\_MIN $= 6$; PIXL rejection threshold
$=$75 (standard for 2--CCD mode); remove hot and flickering
pixels, event grades 0, 2, 3, and 4.  Data from bright and faint
(``bright2'') mode, and high- and medium-bit rate data, were
combined.}
for data screening, extracted the SIS spectra from the
recommended 4$^\prime$ radius region (ABC Guide) and binned to
give $>$10~counts per bin. There were no high background
intervals in the SIS full--field light curves.

Soft (0.5--2 keV) and hard (3--8 keV) SIS X--ray light curves
(Figure 1) in 8 bins, one per orbit, spanning a 40~ks time base,
for the combined SIS0 and SIS1 data, detect NGC~7582 at
$>4\sigma$ in each orbit in both bands.  The hard count rate is
significantly variable ($\chi^2$=21.9), showing a clear
(9.2$\sigma$) 40\% decre<ase in $\sim$6~ks between bins 5
and 6, and remaining low for the rest of the observation
\footnote{In the PKS 2316$-$423 SIS data, we find no
analogous abrupt change in flux; qualitatively, the BL Lac
spectrum is more short--time variable than that of NGC 7582.}
The soft count rate shows no such drop and is consistent
with a constant value ($\chi^2=1.3$), excluding the fifth bin,
which lies $\sim$5$\sigma$ from the mean and may be due to bad
background subtraction.  Including the 5th bin gives $\chi^2 =
4.28$. 
The hardness ratio, which is essentially constant
through bin 5, is significantly lower in bins 6-8.  We refer to
the first five bins as the {\em high state}, and to the remainder
as the {\em low state}.

\bigskip

\centerline{\psfig{figure=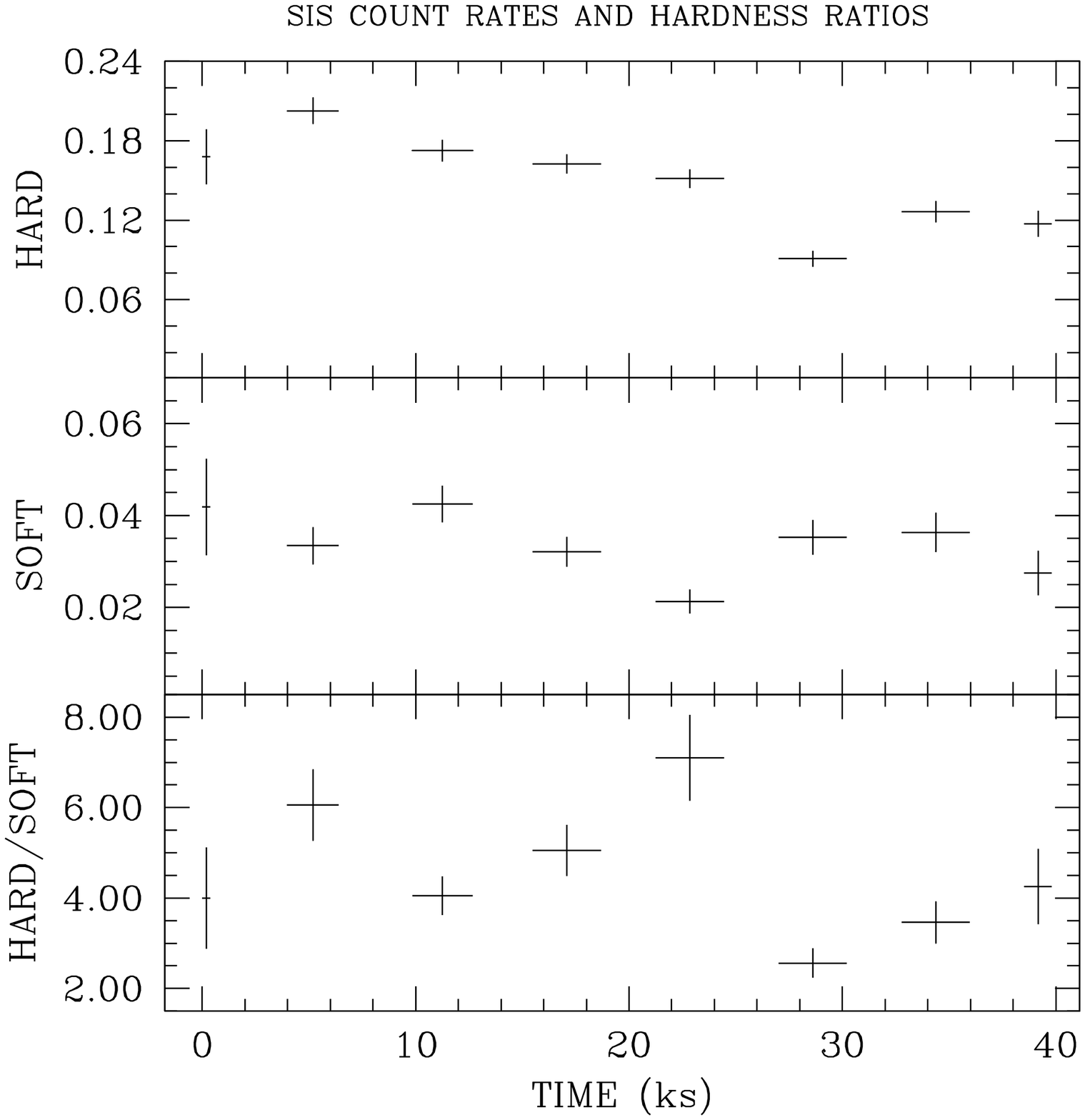,width=3.0in,height=3.0in}}

\figcaption{{\small Background--subtracted 
hard (3-8~keV, top), soft (0.5--2~keV, middle) and
hardness ratio (bottom) light curves for NGC 7582
(SIS0 and SIS1 data combined).} \label{sis_lightcurve}}

\bigskip

We fit the high and low state spectra separately using XSPEC
10.00 (\cite{arn96}), using the 1994 November 9 calibrations. The
Galactic line-of-sight $N_H$ is low ($1.48\times 10^{20}
$~cm$^{-2}$; \cite{elv89}), due to the proximity to the South
Galactic pole.  Both SIS detectors gave consistent results, and
so we report only simultaneous fits (Table~\ref{tbl_fits}). A simple
power--law fit is unacceptable ($\chi^2_\nu > 1.6$; Model A in
Table~\ref{tbl_fits}). 
See Figure 2.
(A fit to just the 2-10 keV range does give an acceptable
power-law fit of $\Gamma = 1.4 \pm 0.3$.)

\bigskip

\centerline{\psfig{figure=fig2.ps,width=2.5in,height=3.0in,angle=-90}}

\figcaption{Best--fit model, data and $\chi^2$ for the simple--power law fit 
(Model A) for simultaneous SIS0 (squares) and SIS1 (triangles) high--state 
observations of NGC 7582.}
\label{spec_fig}

\bigskip

Addition of a low--energy unabsorbed
continuum component, either thermal bremsstrahlung ($kT \sim
0.8$~keV) or a second, steep power-law ($\Gamma\sim 2.5$)
produces a good fit (Models B and C).

\placetable{tbl_fits}

Although a partial covering model (Model D) does significantly
better than a single power law (and the underlying power law
photon index is more typical of canonical AGN values), adding a
soft, unabsorbed component is the preferred model, since an
$F$--test gives a $>$99\% improvement in the fit going from Model
D to Model C.  Interestingly, the partial covering model yields
emission--line--like residuals near 1~keV, as have been reported
in NGC 4151 (\cite{weav94}). No changes in the parameters, except
normalization, are seen (at 90\% confidence) between the two
states. The steep power-law needed at low energies is
inconsistent with the higher energy power-law at high confidence.

Using the power-law plus bremsstrahlung fit (Model B),
the high (low)--state luminosity, L$_X$, of NGC~7582 in units of
10$^{41}$ergs~s$^{-1}$ is 23$^{+6}_{-4}$ (17$^{+7}_{-3}$) in the 2--10~keV
band, and 0.62$^{+0.14}_{-0.23}$ (0.59$^{+0.02}_{-0.05}$) in 
the 0.5--2.0~keV band. So the low
energy, ROSAT band, luminosity is 3\% -- 4\% of the 2-10~keV
luminosity.  (Luminosities are as observed, i.e.  uncorrected for
absorption. We adopt a distance of 27 Mpc [\cite{RCBG3}].)

Absorbing material produces fluorescence lines, notably Fe-K $\alpha$ 
%(because of the high abundance and fluorescent yield of iron).
A narrow ($\sigma$=100~eV) Gaussian line at the cold Fe-K
fluorescence line (6.4~keV), plus a power-law in the energy range
3.5--8 keV, gives 90\% confidence upper limits on the line
equivalent width, $<$230 eV (SIS1, low state), $<$310 eV (SIS0,
high state)
%
%\footnote{The SIS1 high state data are of too low signal-to-noise
%to get a good limit.}
%
. 
The absorbing $N_{H}$ would produce a neutral Fe-K edge (at 7.1
keV) with $\tau_{Fe\,K} = 0.2$. Our 90\% confidence limit is
$\tau_{Fe\,K} \leq 0.2$.

%%%%%%%%%%%%%%%%%%%%%%%%%%%
\subsection{ROSAT HRI Data}  
\label{hri_anal}

The ROSAT HRI observed NGC~7582 for 17.2~ksec on 1995 May 23,
some 6~months after the ASCA observation.  The image
was {\em OBI--shifted} to reduce aspect problems (\cite{mo94}) and
smoothed with a 5$^{\prime\prime}$ Gaussian.  

A source is found positionally coincident with NGC 7582 (using
IRAF/PROS tasks {\tt xexamine, imcnts} and a radius$=$8$^{\prime\prime}$
circle), containing 225~$\pm$~32 cts.  

The source confusion that has dogged NGC~7582 continues with the
discovery of a new bright HRI source 2$^\prime$to the NE (J2000.0:
23$^h$18$^m$29.9$^s$, $-$42$^{\circ}$20$^\prime$
41$^{\prime\prime}$4), which has 90~$\pm$~26~counts, and so 40\% of the
flux of NGC 7582 in the 0.5--2.0 keV band
\footnote{For the NE source, the magnitudes of the two stellar
objects on the digitized sky survey inside the
4.2$^{\prime\prime}$ radius 2$\sigma$ HRI error circle, B=20.2
and 21.1 (\cite{aao97}), allow an AGN or BL~Lac identification
or, for the brighter optical object, a star of type M0V or later
(\cite{js96}).}
. Assuming a $kT=1~$keV bremsstrahlung with Galactic absorption,
this implies a total luminosity (NGC 7582 + NE source) of L$_X=$
$3.9(\pm 0.5)\times$10$^{40}$ ergs~s$^{-1}$.  For the ASCA SIS0
best--fit bremsstrahlung model, the corresponding result is
L$_X=$ $4.9 \times$10$^{40}$ ergs~s$^{-1}$.

The observed full-width half maximum (FWHM) of NGC 7582 (from the IRAF
task {\tt imexamine}) is 4.7$^{\prime\prime}$, and of the NE source
5.0$^{\prime\prime}$. Since the expected FWHM for a point source is
6'' (\cite{Da96}), both NGC 7582 and the NE source are unresolved by
this measure.  However, the HRI image of NGC~7582 appears more
asymmetric than that of the NE source.  To investigate, we divided the
sources into 4 quadrants (\cite{elv83}) of radius 20$^{\prime\prime}$,
with the dividing lines (1) parallel to the N-S and E-W axes; and (2)
rotated by 45$^\circ$.  In both cases, the image of the NE source was
found to be consistent with a symmetric profile, but the image of
NGC~7582 is asymmetric -- brighter on the E side in Case 2 at $>
3~\sigma$, and $>$2.5$\sigma$ fainter on the S side; fainter on the SW
side at $>$2.5$\sigma$, in Case 1.
Subtracting a scaled and shifted image of the NE source from
NGC~7582 gives 2-3~$\sigma$ residuals to the SE, confirming the
asymmetry.  However, given the poor S/N in the NE source
%and the fact that this source is
%elongated in the same direction as NGC 7582 (Fig. 2), 
% [JON: INCONSISTENT WITH EARLIER STATEMENT OF SYMMETRY IN NE!]
we cannot rule out the asymmetry in NGC~7582 being an artifact of
residual ROSAT aspect solution errors.

%%%%%%%%%%%%%%%%%%%%%%%%%%%%%%%%%%%%%%%%%%%%%%%%%%%%%%%%%%%%%%%%%
\section{Discussion}
\label{disc}

An AGN dominates the hard X-ray emission from NGC~7582.  The ASCA
hard X-ray flux drops by 40\% in $\sim$6 ks, implying a source
size of $\sim 2\times 10^{14}$~cm, for which an AGN is the only
known possibility. This is 30 times more variable than
the X-ray flare previously reported by \cite{mush82}.
%which, given the large beam size of the HEAO-A2 experiment, could
%have originated from the BL~Lac PKS~2316$-$423.

The low energy component in the NGC~7582 X-ray spectrum solves
two problems with the \cite{com95} `obscured AGN' model for the
X-ray background: (1) the predicted ROSAT band spectra of NLXGs
should be extremely hard, since the shape will be dominated by
the strongly energy dependent ($E^{-8/3}$) photoelectric
absorption cross-sections (Tucker 1975), yet \cite{rom96} find
quite normal $\Gamma$=1.5 slopes and \cite{cil97} find an even
steeper $\Gamma \sim$2.4; (2) with the predicted N$_H$
distribution, most (70\% -- 90\%) NLXGs should be {\em at least}
10 times brighter in ASCA, yet the four ROSAT NLXGs
studied by
\cite{geo97}, are just 4--12 times brighter in ASCA, implying a
maximum column density of 2.5$\times$10$^{22}$cm$^{-2}$ (for
$\Gamma$=0.8).

Both of these difficulties are removed by an extra, soft, X-ray
spectral feature in NGC~7582: (1) The low energy component has a
power-law slope comparable to that of bright ROSAT NLXGs
(\cite{cil97});
%
%We note that there is still a potential problem
%with the faintest ROSAT NLXGs ($few\times 10^{-15}c.g.s.$) which
%have mean X-ray colors consistent with the X-ray background slope
%(Romero-Colmenero et al., 1996, MNRAS, 282, 94).
%At the faintest fluxes ($few\times 10^{-15} c.g.s.$) though
%misidentifications of NLXGs with random field emission line
%galaxies are possible (Hasinger et al., 1998). The brighter
%($few\times 10^{-14} c.g.s.$) source identifications though seem
%secure.
%
(2) the low energy component prevents the ASCA to ROSAT flux
ratio from exceeding 20, more in line with the ratios implied in
\cite{geo97}. Even so a bright source with this spectrum at
1$\times 10^{-13}$ergs~cm$^{-2}$~s$^{-1}$ in the 2-10~keV band,
would be one of the faintest ROSAT sources in the 0.5-2.0~keV
band, at 3$\times 10^{-15}$ergs~cm$^{-2}$~s$^{-1}$.  Without
the extra soft component, a simple absorbed power-law spectrum
would have a flux of $ 2x10^{-16}$ergs~cm$^{-2}$~s$^{-1}$  in ROSAT,
and so would be undetectable.  The
\cite{geo97} ASCA/ROSAT ratios for NLXGs indicate somewhat larger
uncovered fractions, $\sim$15\%.
%giving a ROSAT flux in the brighter range covered by \cite{boy95}.

What is this soft X-ray component in the ASCA spectrum?  The
constancy of the soft X-rays rules out a simple uncovered fraction
of the central continuum, and instead (with the possible HRI
asymmetries) suggests an extended origin. (Although the hint of
soft-band variability should be borne in mind.) Certainly,
kiloparsec scale extended soft X--ray emission is known in other
AGN (\cite{e90}, \cite{wi92}, \cite{mo95}), where it may be of
nuclear or starburst origin (\cite{wi96}). Hot gas is the likely
dominant emitter in either case.
In NGC~7582, emission line ratios (\cite{wa80}; \cite{bpt}) and
widths (\cite{wh92}) allow either a type2 Seyfert or an
H~II~region dominated object. \cite{wi88} concludes that the
extended `diffuse' radio emission, the 4$^{\prime\prime}$ radius
(1$^{\prime\prime}=130$~pc at 27~Mpc) H$\alpha$ disk
(\cite{mor85}), and the IRAS far-infrared emission in NGC~7582 are
all powered by stellar processes, not by an AGN. A starburst/AGN
composite seems to be required (\cite{mor85};
\cite{wi88}), especially given the hard X-ray variability.

The X-ray luminosity of the soft component in NGC~7582 is modest,
10--100 times smaller than that of
the brighter, reliably identified
(ha98) NLXGs from the "CRSS"
(\cite{cil97}, \cite{rom96}) and comparable with starburst
galaxies (\cite{fab89}). In order to
explain NLXG luminosities up to 10$^{43}$erg~s$^{-1}$, and to
maintain relatively constant soft/hard band X-ray luminosities,
it is more plausible that the soft component is a direct result
of nuclear activity, rather than indirect via a starburst
(\cite{fbai98}).
This favors 
Compton-scattered nuclear emission, or jet/outflow-related emission over
a starburst. Compton reflection could allow some variability if
the `mirror' is close to the nuclear source.

The column density within NGC~7582
decreased by a factor $\sim$6 in the 6 years prior to the ASCA
observations.  The best non-imaging X-ray spectrum of NGC~7582
was obtained with Ginga in 1988 October; \cite{wa93} found $N_H
\approx 50 \times 10^{22}$~cm$^{-2}$, keeping $\Gamma$ fixed at
1.7.  We find, in the ASCA energy range and assuming the same
fixed $\Gamma$, 
$N_H = (7.1^{+0.8}_{-0.4}) \times 10^{22}$~cm$^{-2}$.\footnote{Neither 
Abell~S1101, which was allowed for in the
Warwick et al., analysis, nor PKS 2316$-$423, which was not, has
such a heavily absorbed spectrum (\cite{js98}) }.
A column density at the Ginga level in NLXGs would render the
central AGN component fainter by a factor 10$^{6}$.  This is one
of the larger variations in column density known in AGN, similar
in both timescale and amplitude to those in NGC~4151
(\cite{yaq93}).
Changes on this timescale can be created by motions of molecular
clouds in a starburst model, but the amplitude may be hard to reproduce.

%%%%%%%%%%%%%%%%%%%%%%%%%%%%%%%%%%%%%%%%%%%%%%%%%%%%%%%%%%%%%%%%%
\section{Conclusions}

NGC 7582 is the brightest example of the class of NLXGs found in
the deepest ROSAT observations (\cite{boy95}), with no broad
($>$ 1000 km~s$^{-1}$) components on their optical emission
lines, and may serve as the prototype of the class.  Our
analysis of ASCA and ROSAT observations of NGC~7582 has shown:

\begin{enumerate}
\item A rapid ($\sim$6~ksec) 40\% decrease in the ASCA SIS hard
($>$3~keV) X--ray flux, requiring a compact, AGN source, with an
observed (absorbed) 2--10 keV luminosity of $\sim$ $2 \times
10^{42}$~ergs~s$^{-1}$.
\item An additional, separate, emission component is present in
the ASCA SIS soft band (0.5-2~keV, $L_X\sim 6 \times
10^{40}$~ergs~s$^{-1}$) with a steep ($\Gamma$=2.4) unabsorbed
spectrum.
\item A heavily obscured ($N_H\approx 6\times 10^{22}$~cm$^{-2}$)
nuclear source, that has likely become 6 times less obscured over
the last 6 years.
This may cause difficulty explaining the absorbers as being associated with 
nuclear starbursts, as proposed by \cite{fbai98}.
\end{enumerate}

We also note that the source confusion issues which have dogged
NGC~7582 now seem to be resolved:

\begin{itemize}
\item[4.] NGC~7582 is the dominant source (by a factor of 2--3) of
2--10~keV emission in the field containing itself and the BL Lac
object PKS 2316$-$423.

\item[5.] In the soft X-ray band NGC~7582 dominates the newly
discovered source 2~$^\prime$NE by a factor 2.
\end{itemize}

These observations make it clear that the hard X--ray emission of
NGC~7582, the most `narrow-line' of the NLXGs, is associated with
an AGN. The strong suggestion is that {\em all} NLXG are obscured
AGNs, as hypothesized to explain the hard X-ray background spectral
paradox; but that it is a separate soft X-ray component that
makes NLXGs detectable as ROSAT sources, and provides their
AGN-like X-ray slope. Quite likely this soft component is
extended, and thermal in nature.

%%%%%%%%%%%%%%%%%
\acknowledgments

We thank the anonymous referee for valuable comments that
improved the discussion. We thank K.~Arnaud and G.~Madejski, and
T.~Dotani for help in planning the ASCA observations;
A.~Vikhlinin, D.~Liedahl, G.~Mackie, and K.~Macleod for useful
discussions. This work was supported by NASA grants NAG5--2639
(ASCA), NAG5-1536 and NAG5-2640 (ROSAT), and NASA contract
NAS8--39073 (ASC).

%\newpage
%%%%%%%%%%%%%%%%%%%%%%%%%

%%%%%%%%%%%%%%%%

%%%%%%%
{\small
\begin{table*}
\tablenum{1}
\label{tbl_fits}
\caption{Fits to NGC~7582 Combined ASCA SIS0+SIS1 X-ray Spectra$^a$}
%\begin{center}
\begin{tabular}{lll@{\hspace{0.05in}}lll@{\hspace{0.05in}}l}
&&&&&&\\
\tableline
&&&&&&\\
&\multicolumn{3}{c}{High State$^b$}&\multicolumn{3}{c}{Low State$^b$}\\
Model&$\chi^2/dof$&$N_H^c$
	&$\Gamma$/$k$T,$\Gamma_{unabs}$,$f_{cov}$
	&$\chi^2/dof$&$N_H^c$
	&$\Gamma$/$k$T,$\Gamma_{unabs}$,$f_{cov}$\\
&&&&&&\\
\tableline
&&&&&&\\
A. Single Power-Law (PL)&
233/147& $<$ 0.10&-0.79$^{+0.09}_{-0.11}$&250/146&$<$0.04&-0.63$^{+0.10}_{-0.11}$\\
&&&&&&\\
\tableline
&&&&&&\\
B. PL + Brems.$^d$&
145/144&4.9$^{+2.6}_{-1.2}$&0.61$^{+0.50}_{-0.35}$/(0.81$^{+0.90}_{-0.28}$)&
136/143&5.5$\pm$1.3&0.83$^{+0.43}_{-0.27}$/(0.84$^{+0.46}_{-0.22}$)\\
&&&&&&\\
\tableline
&&&&&&\\
C. Two PLs$^d$&
146/144 &5.3$^{+2.3}_{-1.2}$ &0.64$^{+0.56}_{-0.29}$/(2.5$^{+0.48}_{-0.63}$) &
137/143 &6.0$^{+2.5}_{-1.83}$ &0.87$^{+0.59}_{-0.34}$/(2.4$^{+0.34}_{-0.39}$)\\
&&&&&&\\
\tableline
&&&&&&\\
D. Partially Covered PL&
158/146&7.9$^{+1.3}_{-1.2}$&1.5$^{+0.37}_{-0.40}$/(0.96$^{+0.02}_{-0.03}$)&
152/145&9.1$^{+1.4}_{-1.1}$&1.9$^{+0.32}_{-0.29}$/(0.97$\pm$0.01)\\
&&&&&&\\
\tableline
&&&&&&\\
\end{tabular}
%\end{center}

\medskip

Notes: Photon power law indices are used throughout.
(a) Ranges of parameters are for 90\% confidence level; 
(b) High and low states are as defined in section~\ref{tim_anal};
(c) in units of 10$^{22}$cm$^{-2}$;
(d) bremsstrahlung and second power-law absorption fixed to
N$_H$=N$_H$(Gal). 
\end{table*}
}

%%%%%%%%%%%%%%%
\end{document}